\documentclass[12pt]{emulateapj} 

\usepackage{color}

\DeclareTextSymbol{\degre}{T1}{6}
\DeclareTextSymbol{\degre}{OT1}{23}
\usepackage{wasysym}
\usepackage{array}
\usepackage{graphicx}
\usepackage{float}
\usepackage{hyperref}

\usepackage{textcomp}
\definecolor{red}{rgb}{1.000000,0.000000,0.000000}
\definecolor{IndianRed}{rgb}{0.803922,0.360784,0.360784}

\shorttitle{Carbon--rich planet formation in a solar composition disk}
\shortauthors{Ali-Dib et al.}

\begin{document}


\title{Carbon--rich planet formation in a solar composition disk}

\author{Mohamad Ali-Dib,\altaffilmark{1} Olivier Mousis,\altaffilmark{1} Jean-Marc Petit,\altaffilmark{1} and Jonathan I. Lunine\altaffilmark{2}}

\email{mdib@obs-besancon.fr}

\altaffiltext{1}{Universit{\'e} de Franche-Comt{\'e}, Institut UTINAM, CNRS/INSU, UMR 6213, Observatoire de Besan\c con, BP 1615, 25010 Besan\c con Cedex, France}
\altaffiltext{2}{Center for Radiophysics and Space Research, Space Sciences Building, Cornell University, Ithaca, NY 14853, USA}

\begin{abstract}
The C--to--O ratio is a crucial determinant of the chemical properties of planets. The recent observation of WASP 12b, a giant planet with a C/O value larger than that estimated for its host star, poses a conundrum for understanding the origin of this elemental ratio in any given planetary system. In this paper, we propose a mechanism for enhancing the value of C/O in the disk through the transport and distribution of volatiles. We construct a model that computes the abundances of major C and O bearing volatiles under the influence of gas drag, sublimation, vapor diffusion, condensation and coagulation in a multi--iceline 1+1D protoplanetary disk. We find a gradual depletion in water and carbon monoxide vapors inside the water's iceline with carbon monoxide depleting slower than water. This effect increases the gaseous C/O and decreases the C/H ratio in this region to values similar to those found in WASP 12b's day side atmosphere. Giant planets whose envelopes were accreted inside the water's iceline should then display C/O values larger than those of their parent stars, making them members of the class of so-called ``carbon-rich planets''.
\end{abstract}

\keywords{planets and satellites: atmospheres ---  planets and satellites: composition ---  planets and satellites: formation --- protoplanetary disks}

\section{Introduction}

The C/O ratio is a key parameter for the chemical composition and evolution of giant planets atmospheres since it controls the relative abundances of C-- and O--bearing species. Studies suggest that at equilibrium, as the C/O ratio increases ($\geq$0.8) in the gas phase, all the available O goes into organics, CO, CO$_2$ and CH$_3$OH, so that the gas phase becomes H$_2$O-free and the remaining C is in the form of CH$_4$ \citep{2010ApJ...709.1396F,2011ApJ...743..191M}. The CO/CH$_4$/CO$_2$ ratios in planetary atmospheres are also affected by the possible existence of non equilibrium chemistry effects due to dynamical mixing and photochemistry \citep{2012A&A...546A..43V}. The C/O ratio is also crucial for understanding the chemical evolution of protoplanetary disks \citep{1999JGR...10419003C,1993prpl.conf.1005P}.

Based on comparisons between a grid of model predictions and six narrowband photometry points for the dayside atmosphere of the Hot Jupiter (HJ) WASP 12b, \cite{2011Natur.469...64M} found a best fitting model of C/O$\geq$1 and C/H = 2$\times$10$^{-5}$--1$\times$10$^{-3}$, to compare with the stellar host values of C/O $\sim$ 0.56 and C/H $\sim$ 6$\times$10$^{-4}$. These authors concluded that WASP 12b probably has a superstellar C/O but with a substellar C/H ratio. Because giant planets are expected to have stellar or superstellar metallicities \citep{2005astro.ph..4214K}, new formation models consistent with the properties of WASP 12b have been proposed. For instance, \cite{2011ApJ...743L..16O} considered the H$_2$O, CO and CO$_2$ snowlines in a protoplanetary disk model, and found an increase of the gaseous C/O ratio (compared to the C/O ratio of the stellar host) in the disk regions boxed by these snowlines, even reaching $\sim$1 between the CO and CO$_2$ snowlines. 
They proposed that WASP 12b might have accreted its gaseous envelope in this area but they neglected the dynamical and thermodynamic processes {like gas drag, diffusion and sublimation} that may affect the abundances of volatiles in protoplanetary disks. On the other hand, \cite{2011ApJ...743..191M} have shown that WASP 12b's atmospheric C/O is inconsistent with an atmosphere of the host star's composition, even if enriched in heavy elements by icy planetesimals dissolved in the planet's envelope, and suggest that it may have been formed in an O--depleted part of the protoplanetary disk. 

Here, we present a fully dynamical and multi--snowline model describing the transport and distribution of C and O bearing volatiles in protoplanetary disks, which is inspired from the original idea of \cite{1998Icar..135..537C}. The model is used to compute the C/O ratio as a function of the radial distance in order to constrain the possible formation locations of carbon-rich HJs\footnote{In this manuscript, carbon-rich refers to high C/O ratio, regardless of the C/H value.}. In Sec. 2 we present the different components of our model. Our results are presented in Section 3. In section 4 we discuss the implications and predictions of our model for the formation of carbon-rich HJs, and in particular for WASP 12b.
 
\section{Model prescription}
Here we describe the processes that affect the distribution of volatiles abundances in protoplanetary disks. In our approach, we assume that H$_2$O and CO are the main C--and O--bearing species. We track their respective evolutions near their snowlines, due to multiple effects described below.

\subsection{The protoplanetary disk}
The backbone of our model is the protoplanetary disk model of \cite{2005A&A...442..703H,2006MNRAS.367L..47G}. This is a standard one-dimensional $\alpha$-viscosity disk model, where the viscosity is assumed to be generated by turbulence characterized by the free parameter $\alpha$ according to the formalism of \cite{1973A&A....24..337S}:

\begin{equation}  
\nu = \alpha C_{s}H
\end{equation} 

This model follows \cite{1990AJ.....99..924B} in the (gas+dust) opacity expression and \cite{1991ApJ...375..740R} in the expressions of the isothermal sound speed $C_s$ , and the mid-plane density $\rho_{c}$. The authors also take into account the disk's self--gravity in calculating the scale height $H$. The temperature is calculated by assuming a geometrically thin disk heated by both star illumination and the dissipation of viscous energy. The disk is supposed to be optically thick in the radial direction, but the heat can be transported efficiently vertically. 

{Finally, we chose to use the mean gas radial velocity obtained by \cite{2010ApJ...719.1633H} from their 1+1D $\alpha$-disk model.  In 1+1D and 2D models, in contrast with 1D models, the mean {advective} gas flow experienced by solid particles in the midplane is outward, but the vertically integrated movement is inward as a result of accretion. In all simulations, we used the disk properties obtained from \cite{2005A&A...442..703H} using their inward gas velocity, but imposed the outward velocity from \cite{2010ApJ...719.1633H} in the solid tracking modules only.}

\subsection{The evolution of solids}
\begin{figure}[h]
\begin{center}
\resizebox{\hsize}{!}{\includegraphics[angle=0]{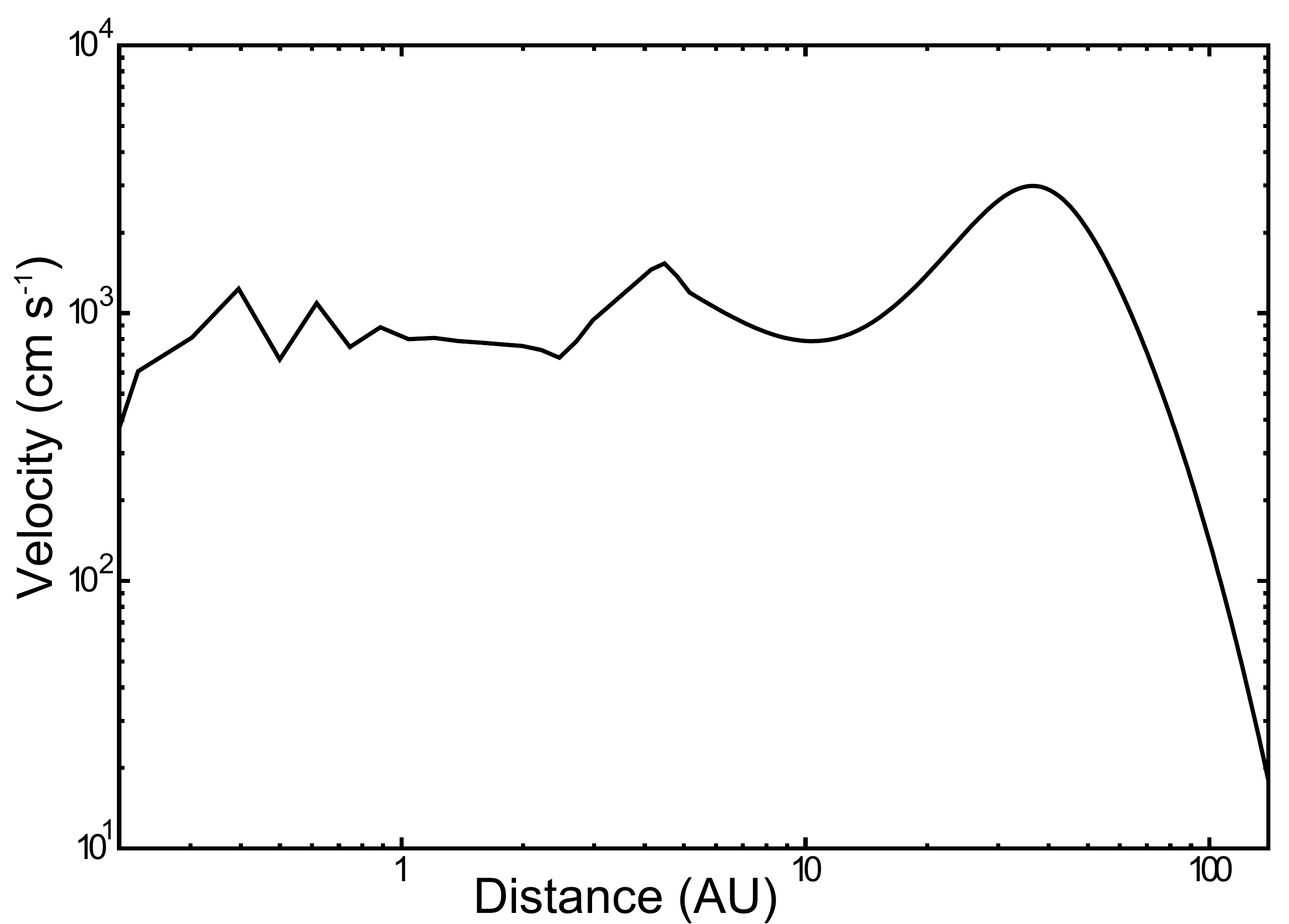}}
\end{center}
\caption{The mean absolute radial velocity of decimetric pebbles as a function of distance. Although this velocity explicitly depends on the particle's density, \textcolor{black}{no significant difference has been found between H$_2$O and CO ices}. Particles placed at 28 AU (CO pebbles) are faster than those at 3.8 AU (H$_2$O) due to \textcolor{black}{the non-linear propagation of gas density in the equations.} The movement direction is always inward. This is equivalent to and reproduces the major aspects of Fig. 2 (third panel) \textcolor{black}{of} \cite{1996A&A...309..301S}.\\}
\label{fig:speed}
\end{figure}

Three important processes affecting solids are incorporated into our model: advection due to solid-gas interaction, sublimation and coagulation. The solids behavior in protoplanetary disks depend strongly on their size \citep{1977MNRAS.180...57W}. \textcolor{black}{Small sub-millimetric particles will remain almost completely coupled to this gas and {advect} outward along with it, although at a slightly lower velocity due to residual gravity. Large particles on the other hand will feel the gas ``headwind'' and get decoupled, drifting inward. This motion is called the gas drag. The exact decoupling size is derived from our model.{ We model this transport using the approach of \cite{1996A&A...309..301S} that incorporates analytically {\color{black} the effects of the disk's turbulence into the gas drag originally predicted by \cite{1977MNRAS.180...57W}. \cite{1996A&A...309..301S} modeled turbulence by dividing the density and velocity of the particles} into mean and fluctuating parts. The {\color{black}gas turbulent velocity is} introduced through the Schmidt number\footnote{{\color{black}The ratio of the viscous to mass diffusion rates.}} used explicitly to express the correlation terms between the mean radial and transverse solids velocities. In the presence of turbulence, the solids will diffuse due to the collective action of turbulent eddies advecting particles in all directions, resulting in additional ``pressure'' and ``viscosity'' terms in the expression {\color{black} determining the mean radial velocity of the solids}. In our model the gas turbulent velocity is calculated self consistently in the \cite{2005A&A...442..703H} {\color{black} framework of the overall disk model}. The relative radial and transverse solids velocities are finally expressed \textcolor{black}{following Eqs. 15 and 16 of} \cite{1996A&A...309..301S}:}} 

\begin{equation} 
2\overline{v_{\phi}} - \frac{\overline{v_{r}}}{\Omega_{k} t_{s*}} = - \frac{1}{\overline{\rho}} \frac{\partial \overline{P}}{\partial r} \frac{1}{\Omega_{k}}
\end{equation}

\begin{eqnarray} 
2 \overline{v_{r}} + \frac{\overline{v_{\phi}}}{\Omega_{k} t_{s*}} = -\overline{V_{r}} - 3r^{-1/2} \frac{\partial}{\partial r}\left(r^{1/2} \frac{\nu}{Sc}\right) \nonumber \\
+ \frac{\left(D - 3\nu\right)}{Sc}\frac{1}{\overline{\rho_{d}}}\frac{\partial \overline{\rho_{d}}}{\partial r} 
\end{eqnarray} 

\noindent where $\overline{P}$ and $\overline{\rho}$ are the gas mean pressure and density, $\Omega_{k}$ is the Keplerian frequency, $\overline{v_{\phi}}$ is the mean relative transverse velocity between particles and gas, $\overline{v_{r}}$ the mean relative radial velocity, $\overline{V_{r}}$ the mean gas radial velocity \textcolor{black}{(from \cite{2010ApJ...719.1633H})}, $t_{s*}$ the stopping time of the particle, $Sc$ the Schmidt number, $D$ the diffusion coefficient, {$r$ the heliocentric distance} and $\overline{\rho_{d}}$ the global {solid particles} density. For numerical simplicity we follow \cite{1996A&A...309..301S} in choosing $D$ = 3$\nu$, a value consistent with more detailed numerical values of $D$ \citep{1988A&A...193..313C}, and which cancels the third term in the second equation. {The presence of a strong vertical magnetic field though might affect the value of $D$} \citep{2006MNRAS.370L..71J}. These equations are coupled to the disk model, and are solved using a globally convergent Newton method \citep{1992nrfa.book.....P}. This module replicated successfully the results of \cite{1996A&A...309..301S} within a small factor probably due to the different disk models used. \textcolor{black}{This module results are presented in Fig. \ref{fig:speed}.} {Our model assumes that the particles have settled in the midplane, where they feel the outward moving gas. A solid particle will settle if its dimensionless Stokes number \footnote{Defined as the ratio between the stopping and eddy turn-over times.} is larger than $\alpha$. \textcolor{black}{ We follow \cite{2010A&A...513A..79B} in expressing this parameter as:}}

\begin{equation}
St = \frac{\rho_s R}{\Sigma_g}\frac{\pi}{2} 
\end{equation}

\noindent {\textcolor{black}{with} $\rho_s$ {\color{black} the single solid} particle's density, $R$ its radius and $\Sigma_g$ the gas surface density. In the case of our decimetric pebbles, \textcolor{black}{we get} $St \sim$ $0.1$ and $0.7$ respectively for H$_2$O and CO particles at \textcolor{black}{the locations of their snowlines}. These values are \textcolor{black}{one} order of magnitude higher than \textcolor{black}{the viscosity parameter used in our calculations ($\alpha = 10^{-2}$), implying that} our assumption of settled particles is valid.}

{The sublimation modeling is based on the simplified approach of \cite{2000Icar..146..525S}. We ignore the condensation of gas on the particle since it is assumed to be minimal in the inner hot regions. We finally write the theoretical maximum sublimation rate of a particle (g cm$^{-2}$ s$^{-1}$) as
\begin{equation}
\chi_{s,max}=-\sqrt{m/(2\pi kT)P_p}
\end{equation}
and derive the particle's radius decrease rate (cm s$^{-1}$) in the form
\begin{equation}
\frac{dR}{dt}=-\frac{0.63\times v_s P_p}{\sqrt{2\times\pi mkT}}
\end{equation}
where \textcolor{black}{$P_{p}$ is the equilibrium vapor pressure on a flat surface for H$_2$O \citep{1992JPhCh..96.8502H}\footnote{P$_p = 1.013 \times 10^6 \exp(-5940/T_p +15.6)$ dyn cm$^{-2}$ } or CO \citep{1985JPCRD..14..849G}\footnote{http://www.nist.gov/data/PDFfiles/jpcrd281.pdf},} $v_s$ is the volume of single molecules, $k$ the Boltzmann constant, {$m$ is the particle's mass and $T$ is its surface temperature}. We multiply the sublimation rate by 0.63 following the experimental results of \cite{1992JPhCh..96.8502H}.}

We model the dust coagulation into larger particles following \cite{2006Icar..181..178C} (Eq. 26): 
\begin{equation}
\Delta\Sigma^{coag} = \Sigma_{d}\left(1 - \exp^{-\Delta t/t_{coag}^r}\right)
\end{equation}
with
\begin{equation}
t_{coag}^r = t_{coag}^{1 AU}\times\frac{\Omega(1 AU)}{\Omega(r)}
\end{equation}

\noindent where $\Sigma_{d}$ is the dust surface density, $\Omega(r)$ is the Keplerian angular rotation velocity at distance $r$ and $t^r_{coag}$ is the average timescale for larger bodies {(in this case decimetric pebbles)} to grow from the local dust supply at distance $r$. We choose t$_{coag}^{1 AU}$ = 10$^{3}$ years, which is the lowest value considered by \cite{2006Icar..181..178C} and others since we do not grow our particles all the way to meter scale in this work. The solids in fact are not allowed to grow larger than 10 cm, since most of the solid mass density in disks is supposed to be in pebbles of this size scale \citep{2012A&A...544A..32L}. We also choose $\Sigma_{d}$ = 1 g/cm$^{-2}$, an average value obtained at $t$ = 10$^{5}$ years in seven runs made by \cite{2006Icar..181..178C}, as the average dust surface density at both snowlines {since we were not able to find any detailed modelling of this quantity in cases relevant to our model}. It is assumed to be constant with respect to time; modeling the expected individual evolutionary trajectories of the surface densities of dust and pebbles is a substantial numerical undertaking beyond the scope of this work.
 
\subsection{The evolution of vapors}
Two processes affecting volatile vapors are included: vapor diffusion and condensation into solid dust.
Any existing volatile vapor is coupled to the H$_2$ gas, {advecting along with it but diffuses according to its own gradient}. The diffusion is modeled following the approach depicted in \cite{1988Icar...75..146S}, \cite{1998Icar..135..537C} and others: 
\begin{equation}
\frac{\delta c}{\delta t} - \frac{2D}{r}\frac{\delta c}{\delta r} -D\frac{\delta^{2}c}{\delta r^{2}} + S(r,t) = 0
\label{eq9}
\end{equation}
 
\noindent where $c$ is the vapor concentration ($c$ = 1 for stellar abundance) and $S(r,t)$ is the source function due to replenishment. This equation is solved and evolved using the explicit Forward time - Centered space (FTCS) method \citep{1992nrfa.book.....P}. 
Once the vapor attains a low enough temperature (160 K for H$_2$O, 25 K for CO), it condenses into solid ice. We assume that the vapor, following the cold finger effect, condenses instantaneously and completely either on existing submillimetric dust or directly as pure ice, leading in both cases to the formation of typical 10$^{-2}$ cm sized dust particles. {This sets a boundary condition for eq. (9) of $c=0$ for the vapor concentration. The particles are then advected outward along with the gas while coagulating into the pebbles that will later replenish the vapor. This concentration of dust at the snowline position might induce an additional diffusive motion. This could result in an extra source term just interior to the snowline. This effect is neglected in the following and will be discussed in section 4.}

{The {\color{black} diffusion of the vapors described} by \textcolor{black}{Eq. \ref{eq9} is key to our results. It describes} the evolution of both \textcolor{black}{H$_2$O and CO vapors} in addition to \textcolor{black}{H$_2$} gas. Any \textcolor{black}{difference} in the behavior of the different gases in the nebula is due to the boundary conditions used to solve this equation. Physically this \textcolor{black}{translates into the presence of snowlines at different locations}. \textcolor{black}{In the presence of turbulent diffusion, the behavior of the various gases is different due to the presence (or absence){\color{black}, and location, } of their snowlines. The flux of the vapor is dominated by the turbulent diffusion.} The presence of a snowline will induce a concentration gradient resulting in an {\color{black}outwardly-directed} vapor flux. In the absence of replenishment, the region inside the snowline will {\color{black} become} depleted on a timescale $t_{diff} = r^2 / D$ where \textcolor{black}{$r$} is the diffusion distance \textcolor{black}{(up to the snowline, if present)}. \textcolor{black}{This implies} different diffusion timescales for H$_2$O and CO \textcolor{black}{and thus relative depletion between the two species}. In the case of the nebular H$_2$ gas for which no snowline is present, $r$ in the above equation is the {\color{black}centrifugal} radius $R_c$\footnote{The radial limit in the disk where the gas stops getting accreted inward and start dispersing outward}. Using \textcolor{black}{Eq. 8 from \cite{2005A&A...442..703H}}, we \textcolor{black}{find $R_c \sim$ 178 AU in the case of our {\color{black}disk} model}\footnote{Using the parameters detailed in \textcolor{black}{Sec. 3}.}}

{ \textcolor{black}{For a CO snowline located at 28 AU, we find $t_{diff}^{H_2} \sim40\times t_{diff}^{CO}$, implying that the H$_2$ diffusive timescale is much longer than that of any condensing species considered in our model. This allows us to use a stationary disk model for the rest of our calculations and decouples the evolution of volatile vapors from that of the nebular gas.} \textcolor{black}{This permits \textcolor{black}{our model} to track directly the concentrations (defined as the ratio of the vapor to $H_2$ gas surface density) instead of the vapor surface density.}

{For a more realistic \textcolor{black}{modelling}, one has to add replenishment to the \textcolor{black}{model} discussed above, {\color{black}as} \textcolor{black}{initially proposed} by \cite{1988Icar...75..146S}. The replenishment is mainly caused by the inward drift and sublimation of the particles created from the condensation of gas into dust at the snowline, and the dust's subsequent coagulation into larger particles. A simple model including this effect was elaborated by \cite{1998Icar..135..537C}. In our model, these replenishing particles are {\color{black}assumed} to be entirely in the form of decimetric pebbles. In the presence of a permanent sink term like a static kilometric planetesimal \citep{2004ApJ...614..490C,2006Icar..181..178C}, the vapor is always depleted no matter what {\color{black}may be} the replenishment speed. In the absence of a permanent sink, the replenishment rate is crucial to the results. Fast replenishment will cause an enrichment of the vapor abundance inside the snowline, while slow replenishment will still result in vapor depletion. In our model the replenishment rate is defined as $S(r,t) = c_{rep}/t_{rep}$ where $c_{rep}$ is the total concentration of the replenishing material present at the snowline position as a function of time, and $t_{rep}$ is the longest time duration between the coagulation and sublimation timescales, \textcolor{black}{which} is {\color{black}the} factor that controls the replenishment speed. We neglect the time an inward pebble needs after its formation is complete in order to reach the snowline since it is too short compared to the other considered timescales.{\color{black} Vapor depletion will only result if $t_{rep}$ is sufficiently long}. The main goal of our model is to calculate accurate replenishment rates to track the concentrations of volatiles inside and at their snowlines.}}

 
\section{Results}
A typical simulation starts with a decimetric pebble (H$_2$O or CO) near its corresponding snowline. This particle is large enough to be decoupled from gas. It will drift inward at the velocity determined by the transport module, and starts sublimating. The time needed for sublimation and the distance traveled before it happens are calculated by the sublimation module. These values are communicated to the vapor diffusion module through the source function. The diffusion module then evolves the vapor concentration inside the snowline. The removed vapor will condense instantaneously at the snowline into 0.1 mm sized dust. This dust, coupled to the gas, will start diffusing outward while coagulating into 10-cm sized pebbles, where it will get decoupled and start drifting inward repeating the cycle.

For the rest of this work, we will use the disk properties at 10$^{5}$ years. {The disk is supposed stationary since {modeling the temporal migration of snowlines require detailed modeling of the accumulated ices present beyond them, which is beyond the scope of this work}. The implication of this assumption on the results should be minimal since the planetary formation timescale is thought to be shorter than the disk evolution timescale \citep{2009Icar..199..338L}.

The initial conditions of the disk model are: $\alpha$ = 0.01 ({same value used by \cite{2010ApJ...719.1633H}}), $M_{cloud}$ = 1 $M_\odot$, $\Omega_{cloud}$ = 3.0$\times$ $10^{-14}$ s$^{-1}$, $T_{cloud}$ = 10 K, $M_{0,star}$ = 0.1 $M_\odot$ and $T_{star}$ = 4000 K. {At $10^5$ years, this leads to a star \textcolor{black}{and} disk system with respective masses of 0.7 and 0.015 $M_{\odot}$, although \textcolor{black}{our system continues to evolve and converges rapidly towards 1 and 0.02 $M_{\odot}$}. The disk properties obtained are typical for protoplanetary disks \citep{2005A&A...442..703H} and give gas surface densities compatible with results reported by \cite{2010ApJ...719.1633H}. The {\color{black}disk surface density, temperature, turbulent velocity and density profiles we use} are given in Fig. \ref{fig:disk}.} Water vapor condenses into crystalline ice at $\sim$ 160 K, and the CO condensation temperature is $\sim$ 25 K \citep{2002crc..book.....L}. This places the H$_2$O and CO snowlines in our disk model at 3.8 AU and 28 AU, respectively. The location of the CO snowline in our model is comparable to that recently inferred at $\sim$ 30 AU in TW Hya \citep{2013Sci...341..630Q}. We assume an initial homogeneous disk C/O ratio of solar value $\sim$ 0.55 \citep{2009ARA&A..47..481A}.

\begin{figure*}[h]
\begin{center}
\resizebox{\hsize}{!}{\includegraphics[angle=0]{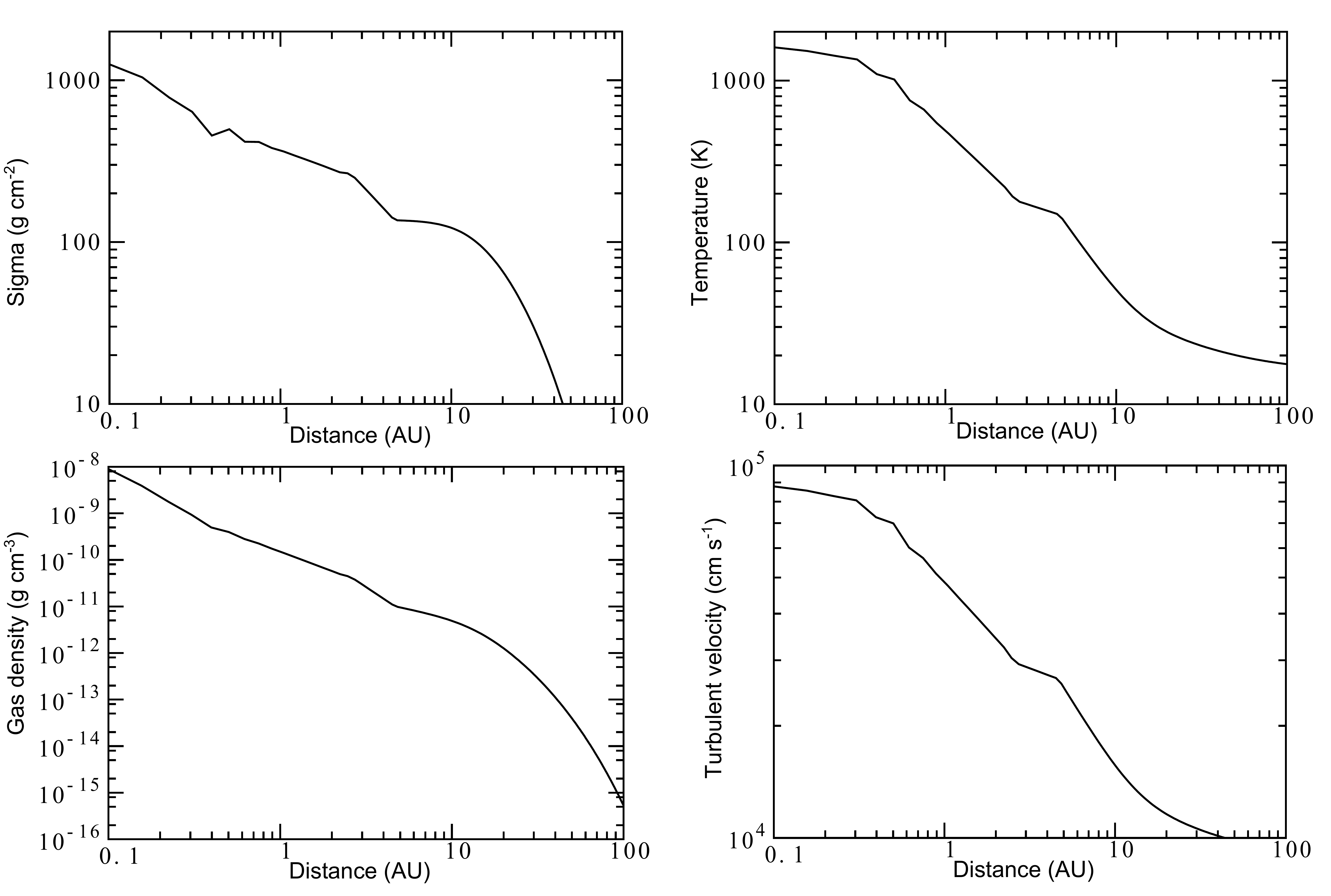}}
\end{center}
\caption{\textcolor{black}{Surface density (top left), temperature (top right), density (bottom left) and turbulent velocity (bottom right) profiles in the midplane of the employed disk model.}}
\label{fig:disk}
\end{figure*}       

Table 1 shows the results from the transport, sublimation and coagulation modules. For H$_2$O solids near the snowline, mm--sized particles and smaller move outward while particles equal to and larger than 1 cm move inward. The decoupling size for H$_2$O particles (where they change direction) is between those two values. 10 cm sized pebbles will drift inward during $\sim$1.0$\times 10^{4}$ years before sublimation, crossing a distance of $\sim$ 2.8 AU. {This contradicts the results of \cite{2006Icar..181..178C} who reported inward traveling distances for particles of no more than a few tenths AU.} This difference might be attributed to the different velocity profiles, disk models and parameters used (\cite{2006Icar..181..178C} used larger $\alpha$ values. Due to the use of an exponential law for depicting its rate, the sublimation occurs almost at the very end of the particle's inward travel. The vapor concentration is thus enhanced in a narrow region both for H$_2$O and CO.
  On the other hand, $10^{-2}$ cm sized dust will diffuse outward and coagulate with other particles for 2.2$\times 10^{4}$ years before reaching pebble-size. Although this seems to corresponds to a traveled distance of $\sim$ 1.6 AU, the real traveled distance should be shorter, since the particle will stop diffusing outward once it reached the decoupling size.

In the case of CO, the decoupling size of particles is between $10^{-2}$ and $10^{-1}$ cm. Decimetric pebbles will move inward during 3.5$\times 10^{3}$ years before sublimating, corresponding to a travel distance of 19.7 AU from their starting position. Coupled $10^{-2}$ cm CO particles also need 2.2$\times 10^{4}$ years in order to coagulate into pebbles, namely the same timescale as for H$_2$O particles, since our approach is species independent and that dust surface densities are assumed equal near both icelines.

{In the case of H$_2$O, \textcolor{black}{the sublimation timescale of decimetric pebbles is close to the coagulation timescale of $10^{-2}$ cm dust (into pebbles), with the {\color{black}latter} slightly longer than the former}. In the case of CO, the {\color{black}pebbles \textcolor{black}{timescale} for sublimation } is much shorter than the dust coagulation timescale. For these reasons, in both cases the replenishment rate is limited by the dust coagulation time ($\sim 2.2\times 10^4$ years), \textcolor{black}{with CO} being more sensitive to this parameter than H$_2$O due to its \textcolor{black}{\color{black}{smaller} } sublimation timescale.}

The evolution of the vapor concentrations is presented in Fig. \ref{fig:main}. In both cases, the vapor diffusion is much faster than replenishment. {The replenishment rate is not enough to counter the effect of diffusion, due to the long coagulation timescales limiting this rate. These long timescales {\color{black} act as a temporary} \textcolor{black}{sink term} that significantly slows down the replenishment.} This leads to a gradual depletion in vapor concentration inside the iceline, {analogous to the depletion reported by \cite{1998Icar..135..537C}, although faster and with lower vapor concentrations due to the different models used}. This depletion is compensated by an increase in solids surface density near the iceline itself. The comparison of the two panels of Fig. \ref{fig:main} shows that the vapor depletion is much faster for H$_2$O than for CO. This is caused by the CO iceline that is much farther out than H$_2$O's iceline, giving CO vapor a longer distance to travel before condensation. {The H$_2$O vapor concentration profile reaches a quasi-depleted stationary state in $2\times10^4$ years. This is faster than the ``no radial drift'' case in \cite{1998Icar..135..537C}, which is mainly due to our closer snowline and higher $D$. {This relatively short diffusive timescale for H$_2$O \textcolor{black}{supports} our assumption of a stationary nebula, since the \textcolor{black}{disk's gas evolves over a 10$^5$--10$^6$ yr timescale} \citep{2005A&A...442..703H}.} Finally, our model was able to reproduce some of the results of \cite{2004ApJ...614..490C} using the correct parameters, and considering {\color{black} meter-sized} bodies instead of pebbles.

\section{Implications for carbon--rich planet formation}
The difference in the timescales needed to deplete water and CO vapors from the region inside the H$_2$O snowline {leads to a gas phase composition compatible with the observed abundances in WASP 12b's dayside atmosphere}. During a long period of time, the CO vapor is the major C and O bearing species in that region, increasing the gas C/O ratio of the area up to $\sim$1 (Fig. \ref{fig:ctoo}). The C/O in this case is never exactly equal to or higher than \textcolor{black}{unity because the residual water vapor slightly decreases the ratio.} Even if the CO vapor \textcolor{black}{exists} in much higher concentrations than H$_2$O in this region, it is still depleted with respect to the initial stellar abundance, leading to a substellar C/H value. For example, Fig. \ref{fig:main} shows that at $t$ = 5.0$\times10^{4}$ years and for $r$ = 2 AU, c$_{H_2O} \sim$ 0.01,{ but c$_{CO} \sim$ 0.55, giving C/O $\sim$ 0.98 and C/H $\sim$ 0.55 times the solar value}. {Any giant planet accreting its envelope from this region in the disk}\textcolor{black}{, assuming no core-envelope coupling, should} have an atmospheric superstellar C/O $\sim$ 1 and a substellar C/H, {corresponding to the properties of the so-called ``carbon rich planets''}. 

A plausible scenario is that WASP 12b's core was formed {through {\color{black}a} streaming instability \citep{2005ApJ...620..459Y}} at or beyond the location of the water snowline where the solids surface density is high (since water vapor is depleted and concentrated as solids at its snowline location in $10^4$ yr), then accreted its atmosphere in the inner region where the C/O ratio is superstellar \textcolor{black}{\citep{2008ApJ...685..560D}}. {This scenario assumes that the core {\color{black} does not mix} with the accreted atmosphere. The opposite case of a well mixed envelope would explain a planet with a higher {\color{black} oxygen but lower carbon abundance} than its host star.} Absent detailed information on the formation location of the planet's core and its subsequent evolution, we can also assume that the entire planet formed in the region inside the water snowline. {Our nominal (first) scenario differs from the one postulated by \cite{2011ApJ...743L..16O}. It has the advantage of ending the formation of WASP 12b in the inner region of the disk, which is much closer to its observed location than  in \textcolor{black}{the scenario of} \cite{2011ApJ...743L..16O}. \textcolor{black}{It is then not necessary} to} invoke long distance migration to explain the observed abundances. It also contradicts their conclusion that superstellar C/O with substellar C/H are compatible only with a planet accreting its atmosphere beyond the water snowline. Further, \textcolor{black}{solid grains do not need a C/O ratio $\sim$0.4 inside the snowline, since according to our dynamical model, H$_2$O ice is quickly depleted.} The same can be said for the CO ice grains inside the corresponding CO snowline.
 
Since the first report of the C/O ratio in WASP 12b dayside atmosphere, it has been the subject of many followup studies and observations. Some of those confirmed the initial inferred abundances \citep{2013MNRAS.435.2268F}, and others disagreed \citep{2013Icar..225..432S}. In the generic volatiles transport model we presented, we made no assumptions regarding WASP 12b. The results are independent from its atmospheric observations. We used a typical protoplanetary disk code and added generic well studied physical effects. Therefore even if future studies definitely ruled out the superstellar C/O ratio in WASP 12b, our results will still be valid and informative since it would help constraint the possible formation location and migration history of planets. Our model is also independent from the absolute positions of the icelines. What is relevant is the relative positions between the two. 

Another important parameter in our model is the coagulation timescale. It controls the CO replenishment rate since it is higher than the sublimation timescale, which is not the case for H$_2$O. For an improbable t$_{coag}^{iceline}$ $\leq$ 3.5$\times 10^{3}$ years, the CO replenishment will become faster than diffusion, allowing for simultaneously superstellar C/H and C/O ratios inside the water's iceline since H$_2$O would still be depleted but CO might be enhanced above solar value if there is enough replenishing matter beyond the snowline. This would result in solids depletion at the snowline and thus excluding this location as a planets forming region. {The growth timescale can also be affected if we take into account the continuous condensation of vapor on dust in addition to the considered coagulation. \cite{2013A&A...552A.137R} found that for $\alpha \sim$ 0.01, the condensation of mm sized dust into 10-cm sized pebbles is possible in 10$^{3}$ $\Omega_k^{-1}$, corresponding in our model to $10^3$ and $10^4$ years respectively near the water and CO snowlines. These timescales would lead to similar results as inferred from our model.}
It is worth mentioning also that in our model we made the implicit assumption that CO is not trapped in water ice clathrates. {This should not affect our main results though since the clathration temperature ($\sim$ 45 K \citep{1985ApJS...58..493L}) would act as a CO snowline, although a more complete model that includes clathrates transport can give important informations in this regard.} In all cases, the water vapor will be depleted much quicker than CO, and the inferred C/O ratio will hold. The C/H ratio on the other hand might be affected by such processes. Finally the observation of a CO snowline by \cite{2013Sci...341..630Q} hints in favor of CO behaving similar to water.

{An important approximation made in our model is to neglect the effect of ices
drifting in from beyond the snowlines on the replenishement rates. We do
account for the inward drift of pebbles, with some delay due to coagulation
time and and time to return to the snowline. Further more, the sublimation of
pebbles is a rather slow process as shown in Table 1 and act as a source a few
AUs inside the snowline. What we are not considering is the return of
microscopic dust grains inside the snowline.}

{Since we assume the condensation or adsorption of vapor is an instantaneous
phenomenon at this scale, then sublimation of any returning grain would also be
instantaneous. As shown in Section 2, the main gas motion in the midplane is a
strong outward advection which carries the miscroscopic grains. The
condensation of vapor at the snowline creates a dust concentration gradient
peaking at this position.  This gradient will cause an isotropic diffusion of
the dust, inducing an inward movement for some fraction of the microscopic
grains. These grains will immediately sublimate as soon as they reach interior
of the snowline, effectively counteracting the vapor gradient imposed by the
cold finger effect. This effect slows down the depletion of the vapors. In
our calculations we supposed the dominance of outward advection which will
substantially mute the inward diffusion due to the much lower advection
timescale and the non-null condensation time in realistic situations.}

{Even considering only the diffusive motion, pebbles will eventually grow from
the grains \citep{2013A&A...552A.137R} and finally allow the formation of a planetary
core at the water snowline. This core will act as a permanent sink rapidly
accreting the newly condensed dust \citep{2004ApJ...614..490C,2006Icar..181..178C},
allowing water to be depleted in 10$^4$ yr and our scenario to hold. This core
will then migrate inward and accrete its envelope from the C/O enhanced gas.}

With current technology, only alkali metals, H$_2$O, CO, CO$_2$ and CH$_4$ are observable in exoplanets atmospheres \citep{2013Icar..226.1654T}. In the near future, with the arrival of {\color{black} JWST, E-ELT and other new astronomical facilities}, additional species should be detectable, so it might be useful to use our model to predict the abundances of other elements. {N should be among the most abundant observable metals in giant planets atmospheres after O and C. In protoplanetary disks, N$_2$ (a volatile) is assumed to be the major N bearing species \citep{1993prpl.conf.1005P}. If we assume that WASP 12b accreted its atmosphere inside the water snowline, N should be accreted in gaseous form leading to a stellar or slightly substellar (due to vapor depletion) NH$_3$ abundance in the deep atmosphere at equilibrium. Since all volatiles are accreted in gaseous form in the envelope, we exclude a possible N enrichment induced by its trapping in clathrate hydrates in the nebula.}


\begin{table*}[h]
\begin{center}
\caption[]{Velocity, evolution time and distance for chosen solids sizes as found by our model}
\begin{tabular}{lccccccc}
\hline
\hline
\noalign{\smallskip}
			&	\multicolumn{3}{c}{H$_2$O}					&			\multicolumn{3}{c}{CO} 													\\
Size (cm) 		& Velocity (cm/s) 			& $\Delta t$ (years) 			& $\Delta R$ (AU) 		& Velocity (cm/s) 		& $\Delta t$ (years) 			& $\Delta R$  (AU) 		\\
\hline
10$^{-2}$ 		& +38.5 					& 2.2$\times 10^{4}$		& +1.78 					& +34.8 				& 2.2$\times 10^{4}$		& +1.61 					\\
10$^{-1}$ 		& +25.3 					& 2.2$\times 10^{4}$		& +1.17 					& -11.0 				& 5.9$\times 10^{5}$		& -14.1 				\\
1 			& -106.0 					& 1.1$\times 10^{4}$		& -2.4 				& -469.6 				& 1.8$\times 10^{4}$		& -18.0 				\\
10 			& -1370.3 					& 9.9$\times 10^{3}$		& -2.8 				& -2647.5 				& 3.5$\times 10^{3}$		& -19.7 				\\
\hline
\end{tabular}\\

Positive and negative velocities mean outward and inward drifts, respectively. $\Delta t$ is the time taken by 1--10 cm particles to drift from their starting positions until sublimation and for 10$^{-2}$--10$^{-1}$ cm gas-coupled dust to coagulate into pebbles {(in this case $\Delta t \equiv t_{coag}^r$)}. $\Delta R$ is the distance travelled by inward drifting particles from their iceline to their sublimation location. The particles are placed initially on their \textcolor{black}{iceline}.
\end{center}
\label{tab:1}
\end{table*}

\begin{figure*}[h]
\begin{center}
\resizebox{\hsize}{!}{\includegraphics[angle=0]{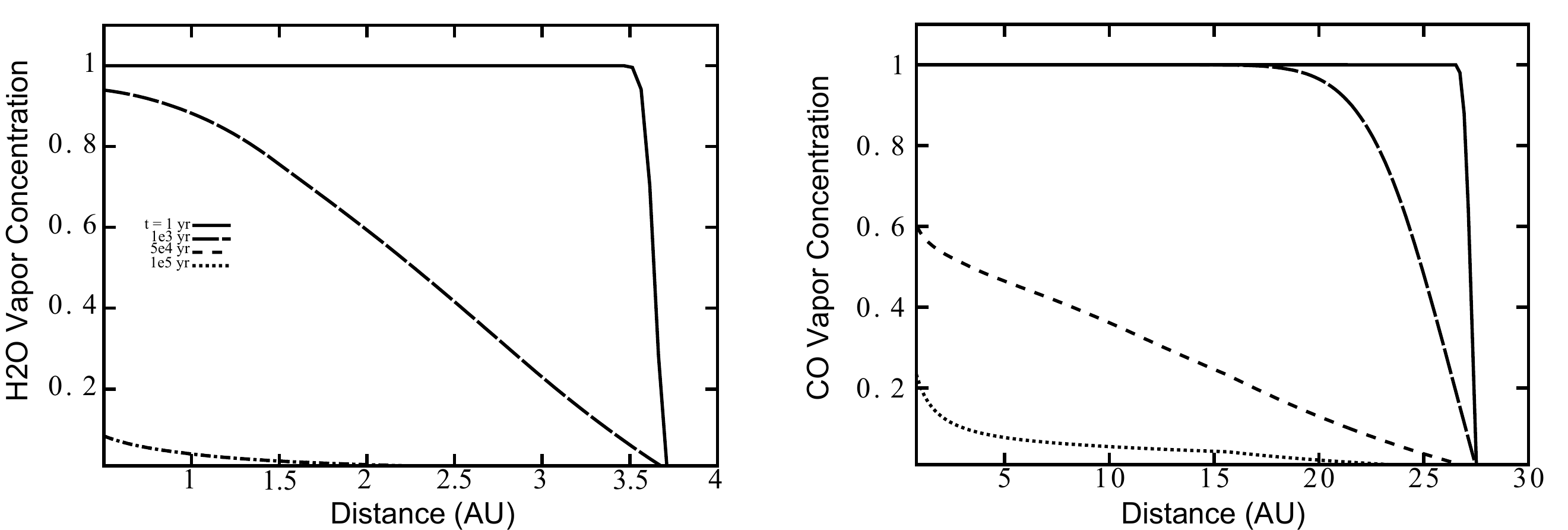}}
\end{center}
\caption{The vapors concentrations of H$_2$O (left panel) and CO (right panel) inside their respective icelines as a function of time and distance to the star. In both cases there is a gradual depletion in the concentration due to diffusion being faster than replenishment. CO depletion is much slower than H$_2$O, leading to a CO--dominated inner nebula. H$_2$O is depleted in $10^4$ yr (not shown), and CO in $10^5$ yr.}
\label{fig:main}
\end{figure*}     

\begin{figure}[h]
\begin{center}
\resizebox{\hsize}{!}{\includegraphics[angle=0]{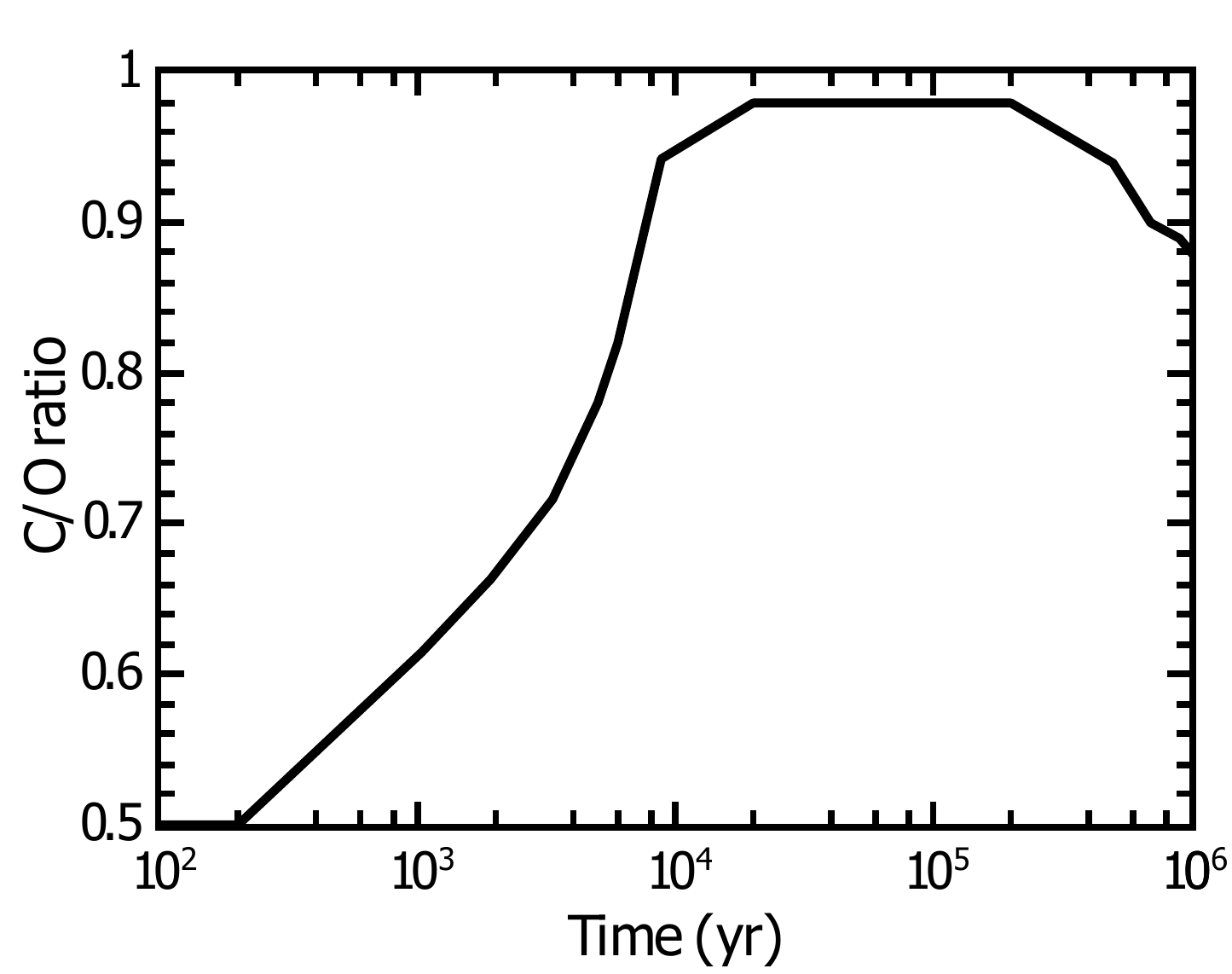}}
\end{center}
\caption{The C/O ratio at 2 AU as a function of time. Early in the disk, \textcolor{black}{H$_2$O vapor depletes much faster than CO vapor does, leading to a progressive increase of the C/O ratio with time}. When \textcolor{black}{the H$_2$O concentration reaches its minimal value, the CO concentration starts to decrease slowly, thus decreasing the C/O ratio in the gas phase}}
\label{fig:ctoo}
\end{figure}

\acknowledgements
We thank an anonymous Referee and \textcolor{black}{S. J. Weidenschilling} for their useful comments that helped us improving our manuscript. We thank T. Guillot and R. Hueso for having provided us with their accretion disk model. Special thanks go to K. Ros for discussions on condensation timescales. MAD is supported by a grant from the city of Besan\c{c}on. O. M. acknowledges support from CNES. JIL acknowledges support from the JWST program through a grant from NASA Goddard.

\end{document}